\documentclass[prd,showpacs,preprintnumbers,amsmath,amssymb,superscriptaddress,floatfix,nofootinbib]{revtex4}
\usepackage{graphicx}
\usepackage{amsmath}
\usepackage{amsfonts}
\usepackage{amssymb}
\usepackage{color}
\usepackage{multirow}
\usepackage{footnote}
\usepackage{ulem}
\usepackage[colorlinks, citecolor=blue,anchorcolor=red,menucolor=red,linkcolor=red,filecolor=red,runcolor=red,urlcolor=blue,frenchlinks=red]{hyperref}

\newcommand{\minv}{M_{\rm inv}}

\newcommand{\mev}{{\rm MeV}}
\newcommand{\tento}[1]{\times 10^{#1}}

\renewcommand\sout{\bgroup \color[rgb]{1,0,0} \ULdepth=-.5ex \ULset}

\begin{document}

\title{$J/\psi \to \gamma \pi\pi$, $\gamma\pi^0\eta$ reactions and the $f_0(980)$ and $a_0(980)$ resonances}

\author{S. Sakai}
\email{shsakai@itp.ac.cn}
\affiliation{Department of Physics, Guangxi Normal University, Guilin 541004, China}
\affiliation{CAS Key Laboratory of Theoretical Physics, Institute of Theoretical Physics, Chinese Academy of Sciences, Beijing 100190, China}

\author{Wei-Hong Liang}
\email{liangwh@gxnu.edu.cn}
\affiliation{Department of Physics, Guangxi Normal University, Guilin 541004, China}
\affiliation{Guangxi Key Laboratory of Nuclear Physics and Technology, Guangxi Normal University, Guilin 541004, China}

\author{G.~Toledo}
\email{toledo@fisica.unam.mx}
\affiliation{Departamento de F\'{i}sica Te\'{o}rica and IFIC, Centro Mixto Universidad de Valencia - CSIC,
Institutos de Investigaci\'{o}n de Paterna, Aptdo. 22085, 46071 Valencia, Spain}
\affiliation{Instituto de Fisica, Universidad Nacional Autonoma de Mexico, AP20-364, Ciudad de Mexico 01000, Mexico}

\author{E.~Oset}
\email{oset@ific.uv.es}
\affiliation{Department of Physics, Guangxi Normal University, Guilin 541004, China}
\affiliation{Departamento de F\'{i}sica Te\'{o}rica and IFIC, Centro Mixto Universidad de Valencia - CSIC,
Institutos de Investigaci\'{o}n de Paterna, Aptdo. 22085, 46071 Valencia, Spain}

\begin{abstract}
We study the $J/\psi \to \gamma \pi^+ \pi^-$, $\gamma \pi^0 \eta$ reactions from the perspective that they come from the $J/\psi \to \phi(\omega) \pi^+ \pi^-, \rho^0\pi^0 \eta$ reactions, where the $\rho^0$, $\omega$, and $\phi$ get converted into a photon via vector meson dominance.
Using models successfully used previously to study the $J/\psi \to \omega (\phi) \pi\pi$ reactions,
we make determinations of the invariant mass distributions for $\pi^+ \pi^-$ in the regions of the $f_0(500)$, $f_0(980)$, and for $\pi^0 \eta$ in the region of the $a_0(980)$.
The integrated differential widths lead to branching ratios below present upper bounds,
but they are sufficiently large for future check in updated facilities.
\end{abstract}



\maketitle

\section{Introduction}
\label{sec:intro}

The low lying scalar mesons, $f_0(500), f_0(980), a_0(980)$ have been the subject of a long debate \cite{Beveren,Bugg,torngvist,Pelaez,Fariborz,Fazio,Briceno}.
The advent of the chiral unitary approach, where input from chiral Lagrangians \cite{Weinberg,Gasser} is used,
and a unitary scheme in coupled channels is followed \cite{npa,Kaiser,Markushin,Juan},
has brought much light into this issue and these resonances appear as dynamically generated from the $\pi\pi, K\bar K, \pi\eta, \eta\eta$ interaction in coupled channels.
From a different perspective,
the $f_0(980)$ was also claimed to be a $K\bar K$ molecular state in Ref.~\cite{Isgur}.
The success of this picture is reflected in the correct description of plenty of reactions concerning their production and decay \cite{review,newrew}.
The studies of Refs.~\cite{npa,Kaiser,Markushin,Juan} are based on the use of the chiral Lagrangians at lowest order,
but the use of next to leading order potentials renders basically the same results \cite{Oop,Rios,PelaRios,GuoOller}.

In a series of papers analyzing data close to threshold \cite{Hanhart,Baru,Kalash},
the authors conclude that the $f_0(980)$ and $a_0(980)$ are not elementary ($q\bar q$) states.
Ultimately, the piling support for the composite structure of these states should come from the ability of this picture to explain physical reactions and predict new ones.
In this direction a boost to this picture was given by the correct interpretation of the $f_0(980)$ and $a_0(980)$ production in the $\phi \to \gamma \pi^0 \pi^0 (\pi^+\pi^-, \pi^0 \eta)$ reactions \cite{Marco,Roca}.
Another boost to the picture came from the study of the $J/\psi \to \phi (\omega) f_0(980)$ reaction in Ref.~\cite{ulfoller} followed by Ref.~\cite{Palomar},
and more challenging from the right prediction of the isospin forbidden $a_0(980)$ production in the same reaction \cite{Hanhartiso,KubisPela,Thomos},
posteriorly confirmed by the BES Collaboration \cite{BESiso} (see also Ref.~\cite{Luisiso}).

The extension of the $J/\psi \to \phi (\omega) f_0(980)$ reaction to the $J/\psi \to \gamma f_0(980)$
should be equally clarifying concerning the nature of the scalar resonances.
Actually, since the $\gamma$ does not have a given isospin, now the $J/\psi \to \gamma a_0(980)$ reaction is equally allowed and the comparison of the production rates introduces new elements to test productions from this molecular picture of scalar mesons.

Experimentally, there are no data in the Particle Data Group (PDG) \cite{pdg} for these decay modes, and only upper limits exist,
but the same spectra that can be associated to $f_0(500)$ and $f_0(980)$ production are available in Ref.~\cite{bsigma}.
Theoretically, there is already one paper making predictions \cite{XiaoOller}.
The model used is the one of Ref.~\cite{Marco}, where $J/\psi \to K\bar K$ and the photon is emitted from the kaons, together with related contact terms.
The $\phi \to K\bar K$ coupling is substituted by the $J/\psi \to K^+ K^-$, which is taken from experiment.

In the present approach, we take a different point of view.
We rely upon the models of Refs.~\cite{ulfoller,Palomar} for $J/\psi \to \phi (\omega) \pi \pi$,
which proved rather successful to interpret experimental data \cite{ningwa,DM2:1,DM2:2,MARK-III}
and implement vector meson dominance (VMD) with $\phi (\omega) \to \gamma$ conversion \cite{Sakurai,Bando}.
Yet, the model has to be extended to include $J/\psi \to \rho^0 PP \to \gamma PP$ (with $P$ the pseudoscalar meson),
and we relate $J/\psi \to \rho^0 PP$ to $J/\psi \to \phi (\omega) PP$
implementing SU(3) symmetry in the primary production vertex $J/\psi \to VPP$,
assuming the $J/\psi (c\bar c)$ as an SU(3) singlet, in the same way as an $s\bar s$ state is assumed to be an isospin singlet.
In this way, we find direct $\rho^0 \pi^0 \eta$ production and $\rho^0 PP$ with $PP$ in isospin $I=1$, which upon final state interaction produces the $a_0(980)$.
The $f_0(980)$ and $a_0(980)$ are, thus, produced without isospin violation, given the fact that the photon carries no determined isospin.
The rates obtained are below the upper experimental bounds but reachable in future experiments.

\section{Formalism}
\label{sec:form}

\subsection{Primary vector-pseudoscalar-pseudoscalar production}\label{subsec:VPP}
\begin{figure}[t]
 \centering
 \includegraphics[width=10cm]{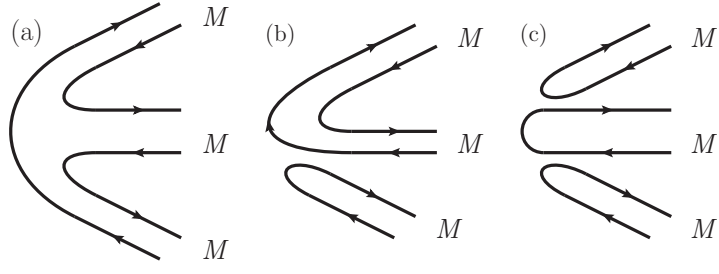}
 \caption{
 Diagrammatic expression of the $\langle MMM \rangle$, $\langle MM \rangle \langle M \rangle$, and $\langle M \rangle\langle M \rangle\langle M \rangle$ terms in Eq.~\eqref{eq:MMM} ($M=P$ or $V$).}
 \label{fig:quark}
\end{figure}

In the $J/\psi \to \phi (\omega) \pi\pi$ reaction studied in Ref.~\cite{ulfoller}, a dominant Okubo-Zweig-Iizuka (OZI) conserving and a subleading OZI violating terms were considered.
An equivalent reformulation of the problem classifying the structures in terms of singlet and octet operators was given in Ref.~\cite{Palomar}
and continued in Ref.~\cite{Lahde}.
In the study of the $J/\psi \to \eta (\eta') h_1(1380)$ reaction done in Ref.~\cite{LiangSakai},
the same primary production vertex was assumed, yet, with a different, more intuitive and practical formulation.
One starts assuming that $J/\psi (c\bar c)$ is a singlet of SU(3), in the same way that an $s\bar s$ state is a singlet of isospin SU(2).
There are then several structures that are SU(3) singlet with two pseudoscalars and one vector,
\begin{equation}
  \langle VPP \rangle,~~\langle V \rangle \, \langle PP \rangle,~~\langle VP \rangle \, \langle P \rangle,~~ \langle V \rangle \, \langle P \rangle^2,\label{eq:MMM}
\end{equation}
with $\langle ... \rangle$ standing for the SU(3) trace,
where $P$ and $V$ stand for the SU(3) pseudoscalar and vector matrices corresponding to $q \bar q$.
This classification was already introduced in the study of the $\chi_{c1} \to \eta \pi^+ \pi^-$ and $\eta_c \to \eta \pi^+ \pi^-$ reactions \cite{LiangXie,LiangVini},
with a primary production of $PPP$.
The $\langle P \rangle^3$ structure was found completely off from data \cite{BESchic1}, and the $\langle PPP \rangle$ was the dominant one.
Consequently, in the study of $J/\psi \to \eta (\eta') h_1(1380)$, the $\langle VPP \rangle$, $\langle V \rangle \langle PP\rangle$ structures were assumed,
with the $\langle VPP \rangle$ being the dominant one,
and a good agreement with data \cite{BESh1} was found.
The formalism was found equivalent to those of Refs.~\cite{ulfoller,Palomar},
and the weight of the $\langle VPP \rangle$ and $\langle V \rangle \langle PP\rangle$
structures were found compatible with the results obtained in Refs.~\cite{ulfoller,Palomar}.
This finding by itself is important since in Refs.~\cite{ulfoller,Palomar} the $PP$ pair was allowed to propagate to generate the $f_0(980)$ state,
while in Ref.~\cite{LiangSakai}, it was a $PV$ pair that was allowed to propagate to generate the $h_1(1380)$ state,
that, within the chiral unitary approach, is dynamically generated from the $VP$ interaction \cite{Lutz,RocaSingh,Gengaxial}.
That both processes are well described starting from the same primary $VPP$ production, allowing either the $PP$ to interact to form the $f_0(980)$,
of the $VP$ to produce the $h_1(1380)$,
speaks much in favor of the dynamically generated nature of these resonances from the meson-meson interaction.
The terms in Eq.~\eqref{eq:MMM} are diagrammatically shown in Fig.~\ref{fig:quark}.
The diagrams (b) and (c), which represent $\langle V \rangle \langle PP \rangle$, $\langle VP \rangle\langle P \rangle$, and $ \langle V \rangle \langle P \rangle^2$, 
have a disconnected part, and the diagram (a) from the $\langle VPP \rangle$ term is preferred in terms of the OZI rule.
The $P$ and $V$ SU(3) matrices corresponding to $q\bar q$ are given by
\begin{equation}\label{eq:phimatrix}
P = \left(
           \begin{array}{ccc}
             \frac{1}{\sqrt{2}}\pi^0 + \frac{1}{\sqrt{3}}\eta + \frac{1}{\sqrt{6}}\eta' & \pi^+ & K^+ \\
             \pi^- & -\frac{1}{\sqrt{2}}\pi^0 + \frac{1}{\sqrt{3}}\eta + \frac{1}{\sqrt{6}}\eta' & K^0 \\
            K^- & \bar{K}^0 & -\frac{1}{\sqrt{3}}\eta + \sqrt{\frac{2}{3}}\eta' \\
           \end{array}
         \right),
\end{equation}
\begin{equation}\label{eq:Vmatrix}
V = \left(
           \begin{array}{ccc}
             \frac{1}{\sqrt{2}}\rho^0 + \frac{1}{\sqrt{2}}\omega  & \rho^+ & K^{*+} \\
             \rho^- & -\frac{1}{\sqrt{2}}\rho^0 + \frac{1}{\sqrt{2}}\omega  & K^{*0} \\
            K^{*-} & \bar{K}^{*0} & \phi \\
           \end{array}
         \right),
\end{equation}
where in $P$ the $\eta$-$\eta'$ mixing of Ref.~\cite{Bramon} has been assumed.

The terms going into $\langle VPP \rangle$ are given below.
\begin{eqnarray}\label{eq:PPV11}
(VPP)_{11} & = & \left( \frac{\rho^0}{\sqrt{2}} + \frac{\omega}{\sqrt{2}} \right)\;
  \left[
    \left( \frac{\pi^0}{\sqrt{2}} + \frac{\eta}{\sqrt{3}} + \frac{\eta'}{\sqrt{6}} \right)^2 + \pi^+ \pi^- + K^+ K^-
  \right] \nonumber \\
&  & + \rho^- \left( \frac{2}{\sqrt{3}}\; \eta \pi^+ +  \frac{2}{\sqrt{6}} \;\eta' \pi^+ + K^+ \bar K^0 \right) \nonumber \\
&&+ K^{*-} \left( \frac{1}{\sqrt{2}} \; \pi^0 K^+ + \pi^+ K^0 + \sqrt{\frac{3}{2}} \; \eta' K^+ \right),
\end{eqnarray}
\begin{eqnarray}\label{eq:PPV22}
(VPP)_{22} & = & \rho^+ \left( \frac{2}{\sqrt{3}} \; \eta \pi^- +  \frac{2}{\sqrt{6}}\; \eta' \pi^- + K^0 K^- \right) \nonumber \\
&& {+}\left( \frac{-\rho^0}{\sqrt{2}} + \frac{\omega}{\sqrt{2}} \right)\;
  \left[ \pi^- \pi^+ +
    \left( \frac{-\pi^0}{\sqrt{2}} + \frac{\eta}{\sqrt{3}} + \frac{\eta'}{\sqrt{6}} \right)^2 + K^0 \bar K^0
  \right] \nonumber \\
&&+ \bar K^{*0} \left(  \pi^- K^+ -\frac{1}{\sqrt{2}} \; \pi^0 K^0 + \sqrt{\frac{3}{2}}\; \eta' K^0 \right),
\end{eqnarray}
\begin{eqnarray}\label{eq:PPV33}
(VPP)_{33} & = &  K^{*+} \left( \frac{1}{\sqrt{2}} \; K^- \pi^0  + \pi^- \bar K^0 + \sqrt{\frac{3}{2}} \; \eta' K^- \right) \nonumber \\
&& +  K^{*0} \left( K^- \pi^+  -\frac{1}{\sqrt{2}} \; \bar K^0 \pi^0  + \sqrt{\frac{3}{2}}\; \eta' \bar K^0 \right) \nonumber \\
&&+ \phi \left[ K^- K^+ + \bar K^0 K^0 + \left( -\frac{1}{\sqrt{3}}\eta + \sqrt{\frac{2}{3}}\eta' \right)^2 \right].
\end{eqnarray}
From these terms, we select those that have a $\omega, \phi, \rho^0$ to implement VMD and we get the weights, $h_i$, for primary production of one vector and two pseudoscalars,
\begin{align}
    h_{\omega \pi^0\pi^0}    &= 1;      & h_{\omega \pi^+\pi^-}  &= \sqrt{2};  &       h_{\omega \eta\eta}  &=\frac{2}{3};
    &  h_{\omega K^+ K^-} &=\frac{1}{\sqrt{2}};   & h_{\omega K^0\bar K^0}  &=\frac{1}{\sqrt{2}};     & h_{\omega \pi^0\eta}  &=0;        \\
      h_{\phi \pi^0\pi^0} &= 0;  & h_{\phi \pi^+\pi^-}  &= 0;  &   h_{\phi \eta\eta}  &=\frac{\sqrt{2}}{3};
    & h_{\phi K^+ K^-} &=1;   & h_{\phi K^0\bar K^0}  &=1;  & h_{\phi \pi^0\eta}  &=0;       \\
     h_{\rho^0 \pi^0\pi^0} &= 0;  & h_{\rho^0 \pi^+\pi^-}  &= 0;  &   h_{\rho^0 \eta\eta}  &=0;
    & h_{\rho^0 K^+ K^-} &=\frac{1}{\sqrt{2}};   & h_{\rho^0 K^0\bar K^0}  &=-\frac{1}{\sqrt{2}};  & h_{\rho^0 \pi^0\eta}  &=\frac{2}{\sqrt{3}}.
\end{align}
where we have multiplied by $\sqrt{2}$ the weights appearing directly from Eqs.\eqref{eq:PPV11}$-$\eqref{eq:PPV33} for $\pi^0 \pi^0, \eta \eta$ production
because we take into account the factor $2!$ for the production of two identical particles and also use, for convenience, the unitary normalization,
\begin{equation*}
  | \pi^0 \pi^0 \rangle  \to  \frac{1}{\sqrt{2}}\, | \pi^0 \pi^0 \rangle, ~~~~~~  | \eta \eta \rangle  \to  \frac{1}{\sqrt{2}}\, | \eta \eta \rangle.
\end{equation*}

Concerning the $\langle V\rangle \, \langle PP \rangle$ structure, we have
\begin{equation}\label{eq:V:PP}
  (\sqrt{2} \omega +\phi) \, ( \pi^0\pi^0 +\pi^+\pi^- +\pi^- \pi^+ +\eta\eta +2 K^+ K^- +2 K^0 \bar K^0),
\end{equation}
from where we get the weights,
\begin{align}
    h'_{\omega \pi^0\pi^0}    &= 2;      & h'_{\omega \pi^+\pi^-}  &= 2\sqrt{2};  &       h'_{\omega \eta\eta}  &=2;
    &  h'_{\omega K^+ K^-} &=2\sqrt{2};   & h'_{\omega K^0\bar K^0}  &=2\sqrt{2};            \\
      h'_{\phi \pi^0\pi^0} &= \sqrt{2};  & h'_{\phi \pi^+\pi^-}  &= 2;  &   h'_{\phi \eta\eta}  &=\sqrt{2};
    & h'_{\phi K^+ K^-} &=2;   & h'_{\phi K^0\bar K^0}  &=2;
\end{align}
and we have ignored the $\eta'$ terms that do not play any relevant role in $f_0, a_0$ production because of its big mass.

If we use the production vertex,
\begin{equation}\label{eq:vertex1}
  A \,\{ \langle VPP\rangle + \beta\, \langle V \rangle  \langle PP\rangle\},
\end{equation}
the production weights will be
\begin{equation}\label{eq:wgt1}
  A\, \{ h_i + \beta \, h'_i\}.
\end{equation}
In the study of Ref.~\cite{LiangSakai}, two solutions for $A$ and $\beta$ were found,
with one of them preferred
and consistent with Refs.~\cite{ulfoller,Palomar}.
We take this solution here corresponding to
\begin{equation}\label{eq:paremeters}
  A=- \tilde{g}; ~~~~~ \tilde{g}= 0.032;  ~~~~~ \beta =0.0927.
\end{equation}
As to the spin structure, following Refs.~\cite{ulfoller,Palomar}, we assume it to be of the type,
\begin{equation}\label{eq:spin:stru}
  \epsilon_\mu (J/\psi) \, \epsilon^\mu (V), ~~(V=\phi, \omega, \rho^0),
\end{equation}
which was found consistent with the experimental information.

\subsection{Vector meson dominance}

\begin{figure}[b!]
\begin{center}
\includegraphics[scale=0.63]{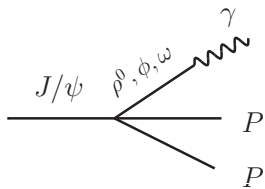}
\end{center}
\vspace{-0.7cm}
\caption{$J/\psi \to \gamma PP$ after $\rho^0,\phi, \omega$ conversion into a photon.}
\label{Fig:1}
\end{figure}

Next, we implement VMD by converting $\rho^0, \omega, \phi$ into a photon.
For this, we use the conversion Lagrangian $V \to \gamma$ \cite{Bando}, and the $J/\psi \to \gamma PP$ starting state is depicted in Fig.~\ref{Fig:1}.

The conversion Lagrangian can be found in a suitable form for the present problem in Ref.~\cite{Nagahiro},
\begin{equation}\label{eq:Lvgamma}
  {\mathcal{L}}_{V\gamma}= -M_V^2 \; \frac{e}{g} \; A_\mu \; \langle V^\mu Q\rangle,
\end{equation}
where 
$A_\mu$ is a photon field,
$e=-|e|,\, \frac{e^2}{4\pi}=\alpha = \frac{1}{137}; \,g= \frac{M_V}{2f_\pi}$ with $M_V = 800$~MeV (a vector-meson
mass) and $f_\pi =93$~MeV.
$Q$ is the quark charge matrix, $Q= \text{diag} (2, -1, -1)/3$.

Considering the $V$ propagator in Fig.~\ref{Fig:1} and the conversion Lagrangian of Eq.~\eqref{eq:Lvgamma},
we find that VMD is implemented with the change,
\begin{equation}\label{eq:change}
  \epsilon_\mu (V) \to \frac{e}{g} \; C_{\gamma V} \, \epsilon_\mu (\gamma),
\end{equation}
where $V$ is any of the $\rho^0, \omega, \phi$ vectors and
\begin{equation}
C_{\gamma V}=
\left\{
             \begin{array}{ll}
             \dfrac{1}{\sqrt{2}}, & \text{for} ~ \rho^0  \\ [0.4cm]
             \dfrac{1}{3\sqrt{2}}, & \text{for} ~ \omega \\  [0.4cm]
             -\dfrac{1}{3}, & \text{for} ~ \phi
             \end{array}
\right.
\end{equation}

\subsection{Final state interaction}
\begin{figure}[b!]
\begin{center}
\includegraphics[scale=0.63]{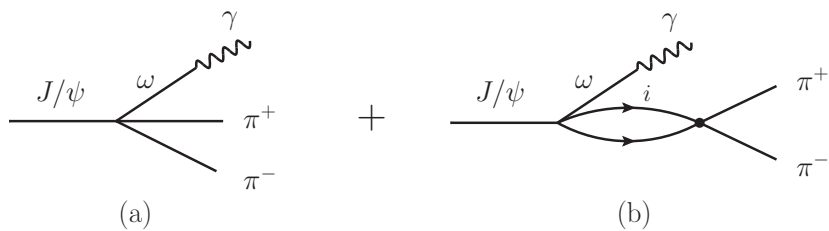}
\end{center}
\vspace{-0.7cm}
\caption{$\pi^+\pi^-$ production driven by $\omega$ conversion. (a) tree level, (b) rescattering. The intermediate states are $i= \pi^0\pi^0, \pi^+\pi^-, K^+ K^-, K^0\bar K^0, \eta \eta$ .}
\label{Fig:2}
\end{figure}

\begin{figure}[b!]
\begin{center}
\includegraphics[scale=0.63]{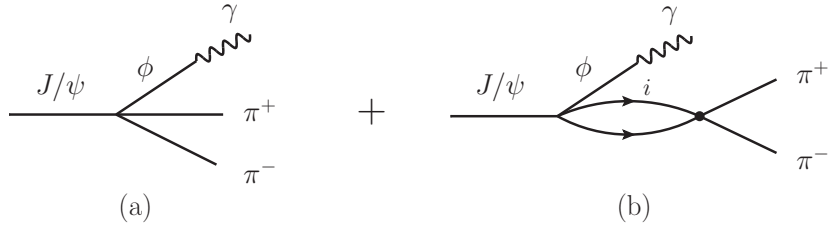}
\end{center}
\vspace{-0.7cm}
\caption{$\pi^+\pi^-$ production driven by $\phi$ conversion. (a) tree level, (b) rescattering. The intermediate states are $i= \eta\eta, K^+ K^-, K^0\bar K^0$ .}
\label{Fig:3}
\end{figure}

\begin{figure}[tbhp]
\begin{center}
\includegraphics[scale=0.63]{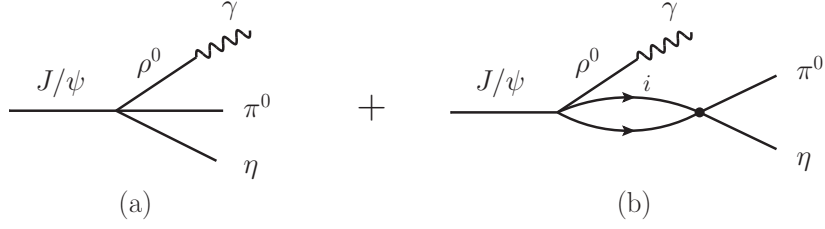}
\end{center}
\vspace{-0.7cm}
\caption{$\pi^0\eta$ production driven by $\rho^0$ conversion. (a) tree level, (b) rescattering. The intermediate states are $i= K^+ K^-, K^0\bar K^0, \pi^0 \eta$ .}
\label{Fig:4}
\end{figure}

Final state interaction is implemented letting the $PP$ pair interact.
This proceeds diagrammatically as shown in Figs.~\ref{Fig:2}, \ref{Fig:3}, and \ref{Fig:4}.

Analytically, we have
\begin{equation}\label{eq:amppipi}
  t_{J/\psi \to \gamma \pi^+\pi^-} = \frac{e}{g} \; \epsilon_\mu (J/\psi) \, \epsilon^\mu (\gamma)\cdot (t_\omega \, C_{\gamma \omega} + t_\phi \, C_{\gamma \phi}),
\end{equation}
with
\begin{equation}\label{eq:tw}
  t_\omega= h_{\omega \pi^+\pi^-} + \sum_i h_{\omega i}\, G_i (M_{\rm inv})\; t_{i, \pi^+\pi^-} (M_{\rm inv}),
\end{equation}
where $M_{\rm inv}$ is the $\pi^+ \pi^-$ invariant mass, $G_i$ are the loop functions of the intermediate states, $i= \pi^0 \pi^0$, $\pi^+\pi^-$, $K^+K^-$, $K^0\bar K^0$, $\eta\eta$, and $t_{i,\pi^+\pi^-}$ the scattering matrices of the chiral unitary approach \cite{npa}.
In the chiral unitary approach, the meson-meson scattering $t$ matrix, $t_{i,j}$, is given by the Bethe-Salpeter equation respecting elastic unitarity with channel coupling,
\begin{align}
 t_{i,j}=&v_{i,j}+v_{i,l}G_lt_{l,j}\notag\\
=&[(1-vG)^{-1}v]_{i,j}.
\end{align}
The interaction kernel $v$ is given by the $s$-wave part of the leading-order chiral Lagrangian (see Refs.~\cite{LiangPLB,DaiPLB} for the explicit form of $v$), 
and the meson-meson loop function $G$,
\begin{align}
 G_i(\sqrt{s})=&i\int\frac{d^4q}{(2\pi)^4}\frac{1}{(P-q)^2-m_{1i}^2+i\epsilon}\frac{1}{q^2-m_{2i}^2+i\epsilon},
\end{align}
where $s=P^2$, 
and $m_{1i}$ and $m_{2i}$ are the meson masses in the channel $i$,
is regularized with a three momentum cutoff $q_{\rm max}$.
In this setup, the $a_0(980)$ peak comes out at the $K\bar{K}$ threshold, and the $f_0(980)$ slightly below this threshold, as a result of the nonperturbative meson-meson interaction.
With respect to Ref.~\cite{npa}, we introduce the $\eta\eta$ coupled channel, and we find that in the cutoff regularization method $q_{\rm max}=600$~MeV is required to fit phenomenology \cite{LiangPLB}.
Similarly,
\begin{equation}\label{eq:tphi}
  t_\phi= h_{\phi \pi^+\pi^-} + \sum_i h_{\phi i}\, G_i (M_{\rm inv})\; t_{i, \pi^+\pi^-} (M_{\rm inv}),
\end{equation}
with $i=\eta\eta, K^+K^-, K^0\bar K^0$.

Analogously, we have
\begin{equation}\label{eq:amppieta}
  t_{J/\psi \to \gamma \pi^0\eta} = \frac{e}{g} \; \epsilon_\mu (J/\psi) \, \epsilon^\mu (\gamma)\; t_{\rho^0} \, C_{\gamma \rho^0},
\end{equation}
with
\begin{equation}\label{eq:trho}
  t_{\rho^0}= h_{\rho^0 \pi^0\eta} + \sum_i h_{\rho^0 i}\, G_i (M_{\rm inv})\; t_{i, \pi^0\eta} (M_{\rm inv}), ~(i= K^+K^-, K^0\bar K^0, \pi^0 \eta).
\end{equation}
The amplitude $t_{i, \pi^0\eta}$ occurs now in $I=1$ and the corresponding matrices,
following Ref.~\cite{npa}, were evaluated in Ref.~\cite{DaiPLB}, again with a cutoff $q_{\rm max}=600$~MeV,
which provides the clear cusp line shape of the $a_0(980)$ as shown in the $\chi_{c1} \to \eta \pi^+ \pi^-$ reaction,
both experimentally \cite{BESchic1} and theoretically \cite{LiangXie}.

\subsection{Consideration of gauge invariance}
Gauge invariance has had an important role in the theoretical study of the $\phi \to \gamma f_0(980) (f_0 \to \pi\pi)$ reaction.
We apply the arguments to $J/\psi \to \gamma f_0(980) (f_0 \to \pi\pi)$ [and the same for $a_0(980)$ production]
following Refs.~\cite{Bramon,Lucio1,Close}.
Taking $P^\mu, K^\mu$ the momenta of the $J/\psi$  and the photon, respectively, the most general structure for the reaction amplitude is given by
\begin{equation}\label{eq:t3}
  t= \epsilon_\mu (J/\psi) \, \epsilon_\nu (\gamma)\; T^{\mu\nu},
\end{equation}
with
\begin{equation}\label{eq:Tmunu}
  T^{\mu\nu} =a \,g^{\mu\nu} +b \, P^\mu P^\nu +c \, P^\mu K^\nu + d \, P^\nu K^\mu +e \, K^\mu K^\nu.
\end{equation}
Gauge invariance ($T^{\mu\nu} K_\nu=0$) forces $b=0$ and requires $d=-a/(P\cdot K)$.
Furthermore, the $c$ and $e$ terms do not contribute to $t$ since $\epsilon_\nu (\gamma)\; K^\nu =0$.
Hence, only the $a$ and $d$ terms are needed, and they are related by the former relationship.
The former arguments are also used in Refs.~\cite{Marco,XiaoOller}, and the $d$ coefficient is explicitly evaluated there.
Here, we have explicitly calculated the $a$ term,
since our amplitudes go as $\epsilon_\mu (J/\psi) \, \epsilon_\nu (\gamma)g^{\mu\nu}$.
We can trivially incorporate the $d$ structure of Eq.~\eqref{eq:Tmunu} into our framework,
but it is unnecessary if we work in the Coulomb gauge [$\epsilon^0(\gamma)=0, ~\vec \epsilon \cdot \vec K =0$],
which explicitly works with transverse photons.
In this gauge, the $d$ term vanishes in the rest frame of the $J/\psi$, $\vec P=0$, since $\vec \epsilon(\gamma) \cdot \vec P =0$.
Thus, we evaluate the decay width in this frame, and we only have to consider the terms from VMD that we have evaluated, with the condition that
\begin{equation}\label{eq:sumpol}
  \sum_{\rm pol.} \epsilon_i(\gamma)\; \epsilon_j(\gamma) = \delta_{ij} - \frac{K_i \, K_j}{\vec K^2}.
\end{equation}
Thus,
\begin{equation}\label{eq:sumsum}
  \overline{\sum} \sum  \epsilon_i(J/\psi)\; \epsilon_i(\gamma) \, \epsilon_j(J/\psi)\; \epsilon_j(\gamma)
  =\frac{1}{3}\, \delta_{ij} \; (\delta_{ij}-\frac{K_i\, K_j}{\vec K^2}) =\frac{2}{3}.
\end{equation}

\section{Results}
We evaluate the differential width for $J/\psi \to \gamma \pi^+\pi^-$,
\begin{align}
 \frac{d\Gamma}{d M_{\rm inv}(\pi^+ \pi^-)}=\frac{1}{(2\pi)^3} \; \frac{A^2}{4M_{J/\psi}^2}\; p_{\gamma}\, \tilde{p}_{\pi}\;\overline{\sum}\sum|t|^2,
 \label{eq:18-1}
\end{align}
where we introduce the normalization factor $A$ of Eq.~\eqref{eq:wgt1} and $p_{\gamma}$ is the photon momentum in the $J/\psi$ rest frame and $\tilde{p}_{\pi}$ the $\pi$ momentum in the $\pi^+\pi^-$ rest frame,
\begin{align}
 p_{\gamma}=&\frac{\lambda^{1/2}(M_{J/\psi}^2,0,\minv^2)}{2M_{J/\psi}},\\
 \tilde{p}_{\pi}=&\frac{\lambda^{1/2}(\minv^2,m_\pi^2,m_\pi^2)}{2\minv},
\end{align}
with $t$ given by Eq.~\eqref{eq:amppipi}, and
\begin{equation}\label{eq:sumt}
  \overline{\sum} \sum |t|^2 = \frac{2}{3} \;  \left( \frac{e}{g} \right)^2 \; |t_\omega \, C_{\gamma \omega} +t_\phi \, C_{\gamma \phi}|^2.
\end{equation}

For $\pi^0 \pi^0$ in the final state, the result is $\frac{1}{2}$ of that of $\pi^+\pi^-$, assuming isospin symmetry, as we have done.

Analogously for $\pi^0 \eta$ in the final state, we have the same differential width substituting $\tilde{p}_{\pi}$ by
\begin{equation*}
  \tilde{p}_{\pi}=\frac{\lambda^{1/2}(\minv^2,m_\pi^2,m_\eta^2)}{2\minv},
\end{equation*}
and $t$ from Eq.~\eqref{eq:amppieta},
\begin{equation}\label{eq:sumt2}
  \overline{\sum} \sum |t|^2 = \frac{2}{3} \;  \left( \frac{e}{g} \right)^2 \; |t_{\rho^0} \, C_{\gamma \rho}|^2.
\end{equation}

Let us recall that when we use the production amplitude of Eq.~\eqref{eq:vertex1}, $\langle VPP\rangle + \beta\, \langle V \rangle  \langle PP\rangle$,
in Eqs.~\eqref{eq:tw}, \eqref{eq:tphi}, and \eqref{eq:trho}, we must substitute $h_i$ by $h_i + \beta h'_i$.
In this case in $t_\phi$ of Eq.~\eqref{eq:tphi}, we must also include $\pi^+\pi^-$, $\pi^0\pi^0$ in the $i$ sum over intermediate states.

\begin{figure}[tbhp]
\begin{center}
\includegraphics[scale=0.85]{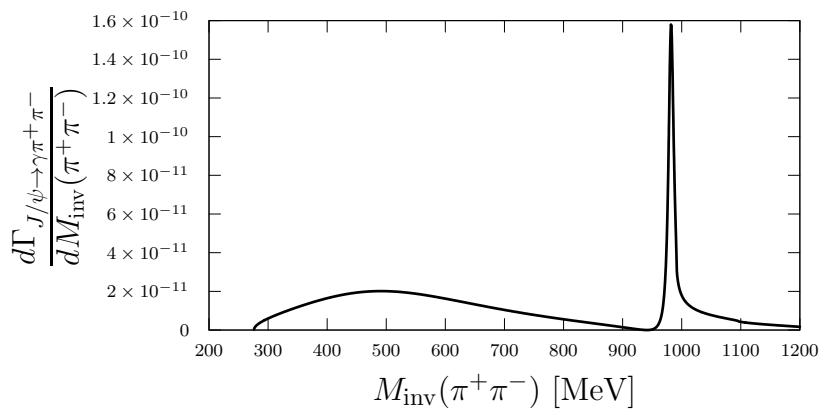}
\end{center}
\vspace{-0.7cm}
\caption{$d\Gamma_{J/\psi\to\gamma\pi^+\pi^-}/dM_{\rm inv}(\pi^+\pi^-)$ as a function of the $\pi^+\pi^-$ invariant mass.}
\label{Fig:5}
\end{figure}

Let us first look at $\pi^+ \pi^-$ production. In Fig.~\ref{Fig:5}, we show $d \Gamma/ d \minv (\pi^+ \pi^-)$ as a function of the $\pi^+\pi^-$ invariant mass.
We find a neat peak for the $f_0(980)$,
but we also find a sizeable strength in the region of the $f_0(500)$.
In Ref.~\cite{XiaoOller}, this region is not investigated, and the $\pi\pi$ invariant mass distribution for $\gamma \pi^+\pi^-$ is plotted from 700~MeV on.
One can envisage a small contribution from the $f_0 (500)$
since the $K\bar K$ channel considered in Ref.~\cite{XiaoOller} couples strongly to $f_0(980)$ but weakly to $f_0(500)$.
Conversely, the $f_0(500)$ couples strongly to $\pi\pi, \eta\eta$ and weakly to $K\bar K$.
In our approach, we have $\pi^+\pi^-$ production in the $f_0(500)$ region through the coefficients $h_{\omega \pi^0 \pi^0}$, $h_{\omega \pi^+ \pi^-}$, $h_{\omega \eta \eta}$, $h'_{\omega \pi^0 \pi^0}$,
$h'_{\omega \pi^+ \pi^-}$, $h'_{\omega \eta \eta}$, $h'_{\phi \pi^0 \pi^0}$, $h_{\phi \pi^+ \pi^-}$ and $h'_{\phi \eta \eta}$,
and we obtain a sizeable contribution of $\pi^+\pi^-$ production in the $f_0(500)$ region.
It is interesting to compare the results of Fig.~\ref{Fig:5} with those of BESIII \cite{bsigma} $0^{++}$ mode of $\pi^0\pi^0$ production.
The results look qualitatively similar.
Although our $f_0(980)$ peak is more prominent than the one in Ref.~\cite{bsigma},
we should note that we would have to take into account the experimental resolution to compare, that would flatten our peak.
The best comparison is the ratio of areas below the peak of the broad $f_0(500)$ region and the $f_0(980)$.
We find a ratio $\Gamma[f_0(500)]/\Gamma[f_0(980)] \simeq 2.8$ versus the experimental one of  $\Gamma[f_0(500)]/\Gamma[f_0(980)] \simeq 5$.
For $\Gamma[f_0(500)]$ and $\Gamma[f_0(980)]$, we integrated $\minv(\pi^+\pi^-)$ in the interval $[2m_\pi,850~\mev]$ and $[950~\mev,1050~\mev]$, respectively.
However, we should note that the fraction of $0^{++}$ around the $f_0(980)$ is very small compared to the total, more than 3 orders of magnitude smaller,
and these numbers should have necessarily large uncertainties.

\begin{figure}[tbp]
\begin{center}
\includegraphics[scale=0.85]{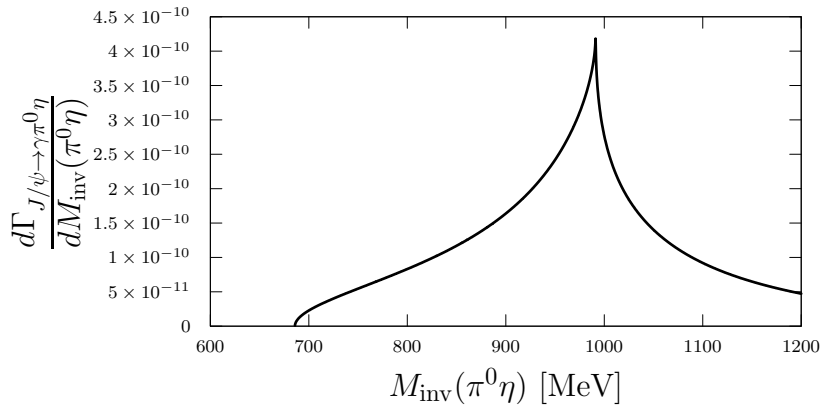}
\end{center}
\vspace{-0.7cm}
\caption{$d\Gamma_{J/\psi\to\gamma\pi^0\eta}/dM_{\rm inv}(\pi^0\eta)$ as a function of the $\pi^0\eta$ invariant mass.}
\label{Fig:6}
\end{figure}

Next we look at the $a_0(980)$ production.
We show in Fig.~\ref{Fig:6} the results for $d \Gamma/ d \minv (\pi^0 \eta)$.
We see a cusplike contribution for the $a_0(980)$ and much strength below the peak.
It is intuitive to think about this latter contribution as coming from the tree level $J/\psi \to \rho^0 \pi^0 \eta \, (\rho^0 \to \gamma)$.
This term does not appear in the framework of Ref.~\cite{XiaoOller}, and one can think of it as responsible for a certain fraction
of the $J/\psi \to \gamma \pi^0 \eta $, non $a_0(980)$, decay reported in Ref.~\cite{BESpieta}.

\begin{figure}[btp]
\begin{center}
\includegraphics[scale=0.85]{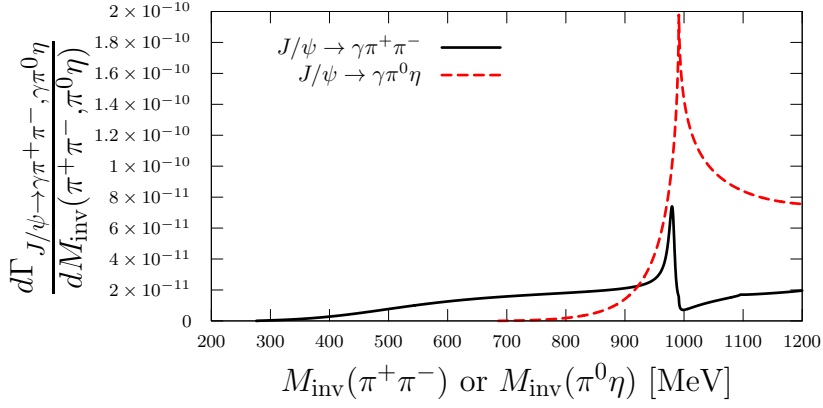}
\end{center}
\vspace{-0.7cm}
\caption{$d\Gamma_{J/\psi\to\gamma\pi^+\pi^-}/dM_{\rm inv}(\pi^+\pi^-)$ and $d\Gamma_{J/\psi\to\gamma\pi^0\eta}/dM_{\rm inv}(\pi^0\eta)$ obtained by eliminating the tree-level contributions.}
\label{Fig:7}
\end{figure}

We next proceed to eliminate the tree level contributions to
obtain what can be better compared with the experimental resonance contribution and with Ref.~\cite{XiaoOller}.
The results are shown in Fig.~\ref{Fig:7}.
We can see that the strength at the peak of the $f_0(980)$ is reduced by about a factor of $2$.
The strength of the $f_0(500)$ does not change much,
but the shape changes appreciably, and the interference effect that made the strength zero around $940~\mev$ is no longer present.
It is interesting to see that a similar interference shows up also in the experimental analysis of Ref.~\cite{bsigma}.
As to the $a_0(980)$, the strength at the peak is also reduced by a factor $2.2$,
but an apparent background from $700~\mev$ up to $950~\mev$ disappears.

\begin{figure}[tbhp]
\begin{center}
\includegraphics[scale=0.85]{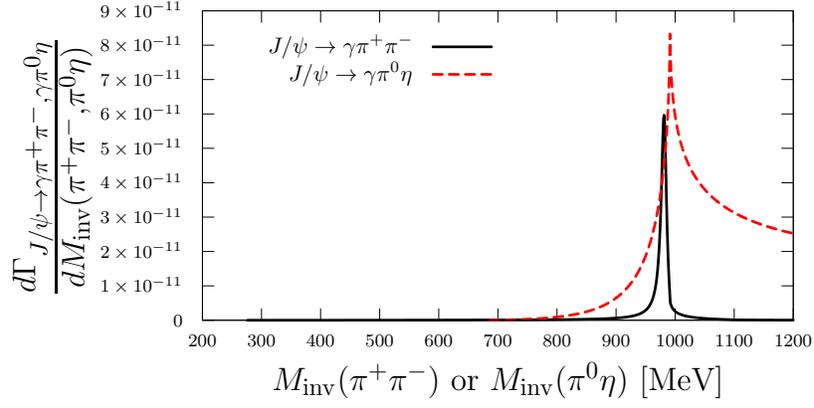}
\end{center}
\vspace{-0.7cm}
\caption{$d\Gamma_{J/\psi\to\gamma\pi^+\pi^-}/dM_{\rm inv}(\pi^+\pi^-)$ and $d\Gamma_{J/\psi\to\gamma\pi^0\eta}/dM_{\rm inv}(\pi^0\eta)$ with $K\bar{K}$ channel only.}
\label{Fig:8}
\end{figure}

To facilitate the comparison with Ref.~\cite{XiaoOller}, driven by $K\bar K$ production,
we also take zero all $h_i$ except $h_{VK\bar K}$.
The results are shown in Fig.~\ref{Fig:8}.
We can see that the peak of the $a_0(980)$ is further reduced by about a factor $2.4$,
and the one of the $f_0(980)$ by about a factor $1.2$.
However, the most striking thing is that the $f_0(500)$ strength disappears totally.
This indicates the relevance of the non $K\bar{K}$ original channels in producing the $f_0(500)$ strength.
This is reminiscent of the picture found in the $B^0 \to J/\psi \pi^+ \pi^-$ and $B^0_s \to J/\psi \pi^+ \pi^-$ reactions \cite{lhcbstone},
where the first reaction shows the clear $f_0(500)$ production and very small $f_0(980)$ production,
while the second one produces clearly the $f_0(980)$ and no sign of the $f_0(500)$.
This was naturally interpreted in Ref.~\cite{LiangPLB},
since after hadronization of a $d\bar{d}$ pair, the first reaction produces mostly $\pi \pi$ but no $K\bar{K}$,
while the second produces mostly $K\bar{K}$  and no $\pi\pi$ after the hadronization of an $s\bar{s}$ pair.

We consider as resonance production our results eliminating tree level and then integrate over $\minv$ [we integrate $\minv$ in the range of $[700~\mev,1200~\mev]$ for $J/\psi\to\gamma a
_0(980)$, $[950~\mev,1050~\mev]$ for $J/\psi\to\gamma f_0(980)$, and $[2m_\pi,850~\mev]$ for $J/\psi\to\gamma f_0(500)$] and dividing by $\Gamma_{J/\psi}$, we find
\begin{align}
 BR [J/\psi \to \gamma a_0(980);~a_0(980)\rightarrow\pi^0\eta]=& 2.7\tento{-7},\\
 BR [J/\psi \to \gamma f_0(980);~f_0(980)\rightarrow \pi^+\pi^-]=& 2.4\tento{-8},\\
 BR [J/\psi \to \gamma f_0(500);~f_0(500)\rightarrow \pi^+\pi^-]=& 6.2\tento{-8}.
\end{align}
We can see that our value of $\Gamma[J/\psi \to \gamma a_0(980)]$ is still smaller than the upper limit of $ BR [J/\psi \to \gamma a_0(980)]=2.5 \times 10^{-6}$.
As to the $\Gamma[J/\psi \to \gamma f_0(980)]$, there is no number in the PDG for it, but as we can see, the $BR$ is quite smaller than that of the $\gamma a_0(980)$.
Comparing with the results of Ref.~\cite{XiaoOller},
$BR(J/\psi\to \gamma a_0(980))=(1.24- 1.61)\tento{-7}$, $BR(J/\psi\to\gamma f_0(980)\to\gamma\pi^+\pi^-)=(0.52- 2.08)\tento{-7}$,
we see that for the $a_0(980)$ production our results are about a factor of 2 bigger than that in Ref.~\cite{XiaoOller}.
The results for $f_0(980)$ production are smaller, 2 times smaller for the lower limit and 8 times smaller for the upper limit.
The orders of magnitude for such small rates, however, agree, providing a fair estimate of the rates
when planning experiments to measure these magnitudes with precision in the future.

We would like to finish this section with some considerations concerning what can one learn from this reaction about the dynamically generated nature of the $f_0(500)$, $f_0(980)$, and $a_0(980)$ resonances.
In principle, one would have to compare with results for the same reaction obtained with different approaches, where the resonances are assumed to be $ q \bar q$ or tetraquarks of no molecular nature.
Such calculations are not available, but it would certainly be useful to have them for comparison.
Yet, in the absence of such results, one can make some interesting observations.
The first one is that the consideration of the resonances as stemming from the interaction of pseudoscalar mesons has allowed us to pin down the mechanism of production and make predictions. This is the most peculiar aspect of the dynamically generated resonances: 
there is not direct production of the resonance in the reaction.
There is direct production of the meson-meson components, and their final state interaction produces the resonances. 
Any other model where the resonances are built up directly from quarks would unavoidable lead to direct production of $J/\psi \to \gamma f_0(500)$, $J/\psi \to \gamma f_0(980)$, or $J/\psi \to \gamma a_0(980)$.
Certainly, the shape of the mass distributions to $\pi \pi$ or $\pi \eta$ would be rather different from those that we have in Figs.~\ref{Fig:5} and \ref{Fig:6}. 
In particular, features like the shape of the $\pi \pi$ mass distribution in Fig.~\ref{Fig:5} are very much tied to the peculiar characteristics of the meson-meson interaction; 
in particular, the zero of the distribution around 950~MeV comes from a subtle cancellation of terms where the phase induced from the $f_0(980)$ relative to the $f_0(500)$ plays an important role, as is the case also in the $\pi \pi$ scattering cross section~\cite{pelaezrep} (see also Sec.~7 of Ref.~\cite{gamgamoller}). 
As, mentioned above, this zero is seen in the $0^{++}$ mode of the $\pi^0 \pi^0$ mass distribution of the BESIII experiment~\cite{bsigma}.

One may argue that the mechanism that we have would also enter into any other picture if instead of generating the $f_0(980)$ and $a_0(980)$ resonances from the rescattering of the meson-meson components, one would replace this amplitude by the pole amplitude obtained from the alternative model. 
Provided that the couplings to the meson-meson components are the same in either approach, the results would be very similar, up to the subtleties about the amplitudes which cannot be parametrized as Breit-Wigner forms, which would be the first choice for the states derived in quark models, for instance. 
These alternative models, made from quark dynamics, would have to be complemented to account for the meson-meson decay, something which is done in certain models~\cite{isgur,davidentem,santopinto}. 
However, as soon as these components are considered, there is automatically a meson-meson component in the wave functions that eats up from the probability of the original quark core. 
The same thing occurs with the $f_0(500)$ as seen in the original papers of Refs.~\cite{Beveren,Bugg,torngvist}.
For a case like the $f_0(980)$ resonance, it is easy to see that the meson probability of $K \bar K$ calculated as $-g^2\partial G/ \partial s$ following the Weinberg compositeness rule~\cite{weinberg} (see Ref.~\cite{danieljuan}) is close to unity \cite{npa,Hanhart,Baru,Kalash}, which renders the resonance as a meson-meson molecule.
Thus, indirectly, one is led again to a molecular state.  
Yet, the insistence of generating the state from the quarks configuration would have as a side effect that the direct production of the resonance would be unavoidable and one would have two mechanisms that interfere, while with the picture we follow, where the resonance is generated from final state interaction of the meson components, there is only one mechanism, which is the one we have evaluated.  
It is thus clear that the alternative pictures would lead to different distributions. 
In that sense, the agreement of an experiment with the predictions done here would give support to the molecular picture that we have described.

Special mention must be made about the $a_0(980)$ state. 
Recent precise experiments show that the $a_0(980)$ appears indeed as a pronounced cusp at the $K \bar K$ threshold~\cite{BESchic1}, as we have obtained in Fig.~\ref{Fig:6}. 
This cusp feature does not come from a bound state, but it is clearly an effect of final state interaction. In our picture, both the  
$f_0(500)$ and $a_0(980)$ structures have the same origin as coming from the interaction of mesons. 
In the case of the $f_0(980)$, the interaction is strong enough to bind the system in the $K \bar K$ channel, and in the case of the $a_0(980)$, the bound-state pole is barely absent
(theoretically, one can get it as a bound state increasing slightly the parameters of the interaction). 
It corresponds to a near bound state and more technically to a virtual state. The appearance of a strong and clean $a_0(980)$ cusp speaks again in favor of the dynamical origin of these resonances. 
Ultimately, the ability of the picture of the dynamical generation to explain many reactions, like the one we have studied, others considered in Refs.~\cite{review,newrew}, and future ones, should serve to pile up support for that picture. 
The predictions made in the present paper should be looked up in this context.

\section{Conclusions}
\label{sec:conc}

We have tackled the $J/\psi \to \gamma \pi^+\pi^-, \gamma \pi^0 \eta$ reactions by taking advantage of previous work on
$J/\psi \to \phi (\omega) \pi\pi (f_0(980))$ and $J/\psi \to \eta (\eta') h_1(1380)$.
These decay modes might look disconnected, but we proved that one can make predictions for one of them using experimental data from the other.
The link stems from the fact that in a first step both reactions proceed creating one vector and two pseudoscalars.
In the first case, the two pseudoscalars interact to produce the $f_0(980)$ state,
and in the second case, a vector and a pseudoscalar interact to produce the axial vector $h_1(1380)$.
The interaction of these pairs of mesons is done using the chiral unitary approach,
and the success of these studies gives further strength to the picture where these resonances are dynamically generated from the interaction of pairs of mesons.

As a next step, we take the common picture of the $VPP$ primary production in $J/\psi \to \phi (\omega) PP$ and $J/\psi \to \eta VP$,
and by means of vector meson dominance, we convert the $J/\psi \to \omega (\phi, \rho^0) PP$ production into $J/\psi \to \gamma PP$ production.
The next step consists in taking into account the $PP$ final state interaction to produce $\pi^+\pi^-, \pi^0 \pi^0$ or $\pi^0 \eta$ at the end,
that produces peaks around the $f_0(500)$ and $f_0(980)$($\pi^0 \pi^0, \pi^+\pi^-$) and around the $a_0(980)$ ($\pi^0 \eta$).
We find a distinct signal for both the $a_0(980)$ and $f_0(980)$ production and a broad distribution in the region of the $f_0(500)$.
The $a_0(980)$ signal is found much larger than that of the $f_0(980)$ but still lower than the experimental upper bound.
Yet, it is not much smaller than this bound, which gives us hope that in future updates these decay modes will be observed.
We also mentioned that the $f_0(980)$ mode was apparently observed, but the small fraction of the total $\gamma \pi^0 \pi^0$ in this mode,
together with the need to separate the different multipole contributions, makes advisable further looks with improved statistics and methods.

Having precise measurements of these decay modes will be an important complement to the $\phi \to \gamma \pi\pi, \gamma \pi^0 \eta$ reactions,
helping us gain insight into the nature of the low mass scalar mesons.

\begin{acknowledgments}
This work is partly supported by the National Natural Science Foundation of China under Grants No.~11565007, No.~11847317, and No.~11975083.
This work is also partly supported by the Spanish Ministerio de Economia y Competitividad
and European FEDER funds under the Contract No.~FIS2011-28853-C02-01, No.~FIS2011-28853-C02-02, No.~FIS2014-57026-REDT, No.~FIS2014-51948-C2-1-P, and No.~FIS2014-51948-C2-2-P.
This work was supported in part by  CONACyT Project No.~252167F and DGAPA-UNAM.
S.~Sakai acknowledges the support by NSFC and DFG through funds provided to the Sino-German CRC110 ``Symmetries and the Emergence of Structure in QCD'' (NSFC Grant No.~11621131001), by the NSFC (Grant No.~11835015, No.~11847612, and No.~11975165), by the CAS Key Research Program of Frontier Sciences (Grant No.~QYZDB-SSW-SYS013) and by the CAS Key Research Program (Grant No.~XDPB09), the CAS Center of Excellence in Particle Physics (CCEPP), the 2019 International Postdoctoral Exchange Program, and the CAS President's International Fellowship Initiative (PIFI) under Grant No.~2019PM0108.
\end{acknowledgments}



\begin{thebibliography}{99}

%

\bibitem{Beveren}
  E.~van Beveren, T.~A.~Rijken, K.~Metzger, C.~Dullemond, G.~Rupp, and J.~E.~Ribeiro,
  Z.\ Phys.\ C {\bf 30}, 615 (1986).

\bibitem{Bugg}
  E.~van Beveren, D.~V.~Bugg, F.~Kleefeld, and G.~Rupp,
  Phys.\ Lett.\ B {\bf 641}, 265 (2006).

\bibitem{torngvist}
F.~E.~Close and N.~A.~Tornqvist,
   J.\ Phys.\ G {\bf 28}, R249 (2002).

\bibitem{Pelaez}
  J.~R.~Pelaez,
  Phys.\ Rev.\ Lett.\  {\bf 92}, 102001 (2004).

\bibitem{Fariborz}
  D.~Black, A.~H.~Fariborz, F.~Sannino, and J.~Schechter,
  Phys.\ Rev.\ D {\bf 59}, 074026 (1999).

\bibitem{Fazio}
  F.~De Fazio and M.~R.~Pennington,
  Phys.\ Lett.\ B {\bf 521}, 15 (2001).

\bibitem{Briceno}
  R.~A.~Briceno, J.~J.~Dudek, R.~G.~Edwards, and D.~J.~Wilson,
  Phys.\ Rev.\ D {\bf 97}, 054513 (2018).

\bibitem{Weinberg}
  S.~Weinberg,
  Phys.\ Rev.\  {\bf 166}, 1568 (1968).

\bibitem{Gasser}
  J.~Gasser and H.~Leutwyler,
  Annals Phys.\  {\bf 158}, 142 (1984).

\bibitem{npa}
  J.~A.~Oller and E.~Oset,
  Nucl.\ Phys.\ A {\bf 620}, 438 (1997);
  Erratum: [Nucl.\ Phys.\ A {\bf 652}, 407 (1999)].


\bibitem{Kaiser}
  N.~Kaiser,
  Eur.\ Phys.\ J.\ A {\bf 3}, 307 (1998).

\bibitem{Markushin}
  M.~P.~Locher, V.~E.~Markushin, and H.~Q.~Zheng,
  Eur.\ Phys.\ J.\ C {\bf 4}, 317 (1998).


\bibitem{Juan}
  J.~Nieves and E.~Ruiz Arriola,
  Nucl.\ Phys.\ A {\bf 679}, 57 (2000).

\bibitem{Isgur}
  J.~D.~Weinstein and N.~Isgur,
  Phys.\ Rev.\ D {\bf 41}, 2236 (1990).


\bibitem{review}
  J.~A.~Oller, E.~Oset, and A.~Ramos,
  Prog.\ Part.\ Nucl.\ Phys.\  {\bf 45}, 157 (2000).


\bibitem{newrew}
  E.~Oset {\it et al.},
  Int.\ J.\ Mod.\ Phys.\ E {\bf 25}, 1630001 (2016).

\bibitem{Oop}
  J.~A.~Oller, E.~Oset, and J.~R.~Pelaez,
  Phys.\ Rev.\ D {\bf 59}, 074001 (1999);
  Erratum: [Phys.\ Rev.\ D {\bf 60}, 099906 (1999)];
  Erratum: [Phys.\ Rev.\ D {\bf 75}, 099903 (2007)].

\bibitem{GuoOller}
  Z.~H.~Guo and J.~A.~Oller,
  Phys.\ Rev.\ D {\bf 84}, 034005 (2011).

\bibitem{PelaRios}
  J.~R.~Pelaez and G.~Rios,
  Phys.\ Rev.\ D {\bf 82}, 114002 (2010).

\bibitem{Rios}
  J.~R.~Pelaez and G.~Rios,
  Phys.\ Rev.\ Lett.\  {\bf 97}, 242002 (2006).

\bibitem{Hanhart}
  V.~Baru, J.~Haidenbauer, C.~Hanhart, Y.~Kalashnikova, and A.~E.~Kudryavtsev,
  Phys.\ Lett.\ B {\bf 586}, 53 (2004).

\bibitem{Baru}
 V.~Baru, J.~Haidenbauer, C.~Hanhart, A.~E.~Kudryavtsev, and U.~G.~Meissner,
  Eur.\ Phys.\ J.\ A {\bf 23}, 523 (2005).


\bibitem{Kalash}
  V.~Baru, C.~Hanhart, Y.~S.~Kalashnikova, A.~E.~Kudryavtsev, and A.~V.~Nefediev,
  Eur.\ Phys.\ J.\ A {\bf 44}, 93 (2010).

\bibitem{Marco}
  E.~Marco, S.~Hirenzaki, E.~Oset, and H.~Toki,
  Phys.\ Lett.\ B {\bf 470}, 20 (1999).


\bibitem{Roca}
  J.~E.~Palomar, L.~Roca, E.~Oset, and M.~J.~Vicente Vacas,
  Nucl.\ Phys.\ A {\bf 729}, 743 (2003).

\bibitem{ulfoller}
  U.~G.~Meissner and J.~A.~Oller,
  Nucl.\ Phys.\ A {\bf 679}, 671 (2001).


\bibitem{Palomar}
  L.~Roca, J.~E.~Palomar, E.~Oset, and H.~C.~Chiang,
   Nucl.\ Phys.\ A {\bf 744}, 127 (2004).

\bibitem{Hanhartiso}
  Y.~S.~Kalashnikova, A.~E.~Kudryavtsev, A.~V.~Nefediev, C.~Hanhart, and J.~Haidenbauer,
  Eur.\ Phys.\ J.\ A {\bf 24}, 437 (2005).

\bibitem{KubisPela}
  C.~Hanhart, B.~Kubis, and J.~R.~Pelaez,
  Phys.\ Rev.\ D {\bf 76}, 074028 (2007).

\bibitem{Thomos}
  T.~Branz, T.~Gutsche, and V.~E.~Lyubovitskij,
  Phys.\ Rev.\ D {\bf 78}, 114004 (2008).

\bibitem{BESiso}
  M.~Ablikim {\it et al.} [BESIII Collaboration],
  Phys.\ Rev.\ D {\bf 83}, 032003 (2011).

\bibitem{Luisiso}
  L.~Roca,
  Phys.\ Rev.\ D {\bf 88}, 014045 (2013).

\bibitem{pdg}
M.~Tanabashi {\it et al.} [Particle Data Group],
   Phys.\ Rev.\ D {\bf 98}, no. 3, 030001 (2018).

\bibitem{bsigma}
  M.~Ablikim {\it et al.} [BESIII Collaboration],
  Phys.\ Rev.\ D {\bf 92},  052003 (2015);
  Erratum: [Phys.\ Rev.\ D {\bf 93}, 039906 (2016)].

\bibitem{XiaoOller}
  C.~W.~Xiao, U.-G.~Mei\ss ner, and J.~A.~Oller,
  arXiv:1907.09072 [hep-ph].


\bibitem{ningwa}
N.~Wu,
  hep-ex/0104050.

\bibitem{DM2:1}
  J.~E.~Augustin {\it et al.} [DM2 Collaboration],
  Nucl.\ Phys.\ B {\bf 320}, 1 (1989).

\bibitem{DM2:2}
  A.~Falvard {\it et al.} [DM2 Collaboration],
  Phys.\ Rev.\ D {\bf 38}, 2706 (1988).

\bibitem{MARK-III}
W.S. Lockman, MARK-III Collaboration, SLAC-PUB-5139, Presented in: 3rd International Conference on Hadron Spectroscopy, Ajaccio, France, 23-27 September, 1989.

\bibitem{Sakurai}
J.J. Sakurai, Currents and Mesons, University of Chicago Press, Chicago, 1969.

\bibitem{Bando}
  M.~Bando, T.~Kugo, S.~Uehara, K.~Yamawaki, and T.~Yanagida,
  Phys.\ Rev.\ Lett.\  {\bf 54}, 1215 (1985).

\bibitem{Lahde}
  T.~A.~Lahde and U.~G.~Meissner,
  Phys.\ Rev.\ D {\bf 74}, 034021 (2006).

\bibitem{LiangSakai}
  W.~H.~Liang, S.~Sakai, and E.~Oset,
  Phys.\ Rev.\ D {\bf 99}, 094020 (2019).

\bibitem{LiangXie}
  W.~H.~Liang, J.~J.~Xie, and E.~Oset,
  Eur.\ Phys.\ J.\ C {\bf 76}, 700 (2016).

\bibitem{LiangVini}
  V.~R.~Debastiani, W.~H.~Liang, J.~J.~Xie, and E.~Oset,
  Phys.\ Lett.\ B {\bf 766}, 59 (2017).

\bibitem{BESchic1}
  M.~Ablikim {\it et al.} [BESIII Collaboration],
  Phys.\ Rev.\ D {\bf 95}, 032002 (2017).

\bibitem{BESh1}
  M.~Ablikim {\it et al.} [BESIII Collaboration],
  Phys.\ Rev.\ D {\bf 98},  072005 (2018).


\bibitem{Lutz}
  M.~F.~M.~Lutz and E.~E.~Kolomeitsev,
  Nucl.\ Phys.\ A {\bf 730}, 392 (2004).

\bibitem{RocaSingh}
  L.~Roca, E.~Oset, and J.~Singh,
  Phys.\ Rev.\ D {\bf 72}, 014002 (2005).

\bibitem{Gengaxial}
  Y.~Zhou, X.~L.~Ren, H.~X.~Chen, and L.~S.~Geng,
  Phys.\ Rev.\ D {\bf 90},  014020 (2014).

\bibitem{Bramon}
  A.~Bramon, A.~Grau, and G.~Pancheri,
  Phys.\ Lett.\ B {\bf 283}, 416 (1992).

\bibitem{Nagahiro}
  H.~Nagahiro, L.~Roca, A.~Hosaka, and E.~Oset,
  Phys.\ Rev.\ D {\bf 79}, 014015 (2009).

\bibitem{LiangPLB}
  W.~H.~Liang and E.~Oset,
  Phys.\ Lett.\ B {\bf 737}, 70 (2014).

\bibitem{DaiPLB}
  J.~J.~Xie, L.~R.~Dai, and E.~Oset,
  Phys.\ Lett.\ B {\bf 742}, 363 (2015).

\bibitem{Lucio1}
  J.~L.~Lucio Martinez and J.~Pestieau,
  Phys.\ Rev.\ D {\bf 42}, 3253 (1990).

\bibitem{Close}
  F.~E.~Close, N.~Isgur, and S.~Kumano,
  Nucl.\ Phys.\ B {\bf 389}, 513 (1993).

\bibitem{BESpieta}
  M.~Ablikim {\it et al.} [BESIII Collaboration],
  Phys.\ Rev.\ D {\bf 94},  072005 (2016).

\bibitem{lhcbstone}
  R.~Aaij {\it et al.} [LHCb Collaboration],
  Phys.\ Lett.\ B {\bf 698} (2011) 115
  [arXiv:1102.0206 [hep-ex]].

\bibitem{pelaezrep} 
  J.~R.~Pelaez,
  Phys.\ Rept.\  {\bf 658}, 1 (2016)

\bibitem{gamgamoller} 
  J.~A.~Oller and E.~Oset,
  Nucl.\ Phys.\ A {\bf 629}, 739 (1998)

\bibitem{isgur} 
  S.~Godfrey and N.~Isgur,
  Phys.\ Rev.\ D {\bf 32}, 189 (1985).

\bibitem{davidentem} 
  J.~Segovia, D.~R.~Entem, and F.~Fernández,
  Phys.\ Lett.\ B {\bf 715}, 322 (2012)

\bibitem{santopinto} 
  H.~García-Tecocoatzi, R.~Bijker, J.~Ferretti, and E.~Santopinto,
  Eur.\ Phys.\ J.\ A {\bf 53}, no. 6, 115 (2017)


\bibitem{weinberg} 
  S.~Weinberg,
  Phys.\ Rev.\  {\bf 137}, B672 (1965).

\bibitem{danieljuan} 
  D.~Gamermann, J.~Nieves, E.~Oset, and E.~Ruiz Arriola,
  Phys.\ Rev.\ D {\bf 81}, 014029 (2010)


\end{thebibliography}
  \end{document}